\documentstyle[12pt,aasms]{article}
\tighten

\slugcomment{Submitted to ApJ}

\begin{document}

\title{On the efficiency of grain alignment in dark clouds}

\author{A. Lazarian\altaffilmark{1,4}, Alyssa A.
Goodman\altaffilmark{2,3},
Philip C. Myers\altaffilmark{4}}

\altaffilmark{1}{Princeton University Observatory, Princeton, NJ 08544}

\altaffilmark{2}{Department of Astronomy, Harvard University, 
60 Garden Street, Cambridge, MA 02138}

\altaffilmark{3}{National Science Foundation Young Investigator}

\altaffilmark{4}{Harvard-Smithsonian Center for Astrophysics,  
60 Garden Street,
Cambridge, MA 02138}

\begin{abstract}
A quantitative analysis of grain alignment in the filamentary dark cloud 
L1755 in Ophiuchus is presented. We show, that the observed decrease of
the 
polarization to extinction ratio for the inner parts of 
 this quiescent dark cloud can be explained as a result of
the 
decrease of the efficiency of grain alignment.
We make quantitative estimates of grain alignment efficiency for six
 mechanisms, involving grains with either thermal or suprathermal
rotation, interacting 
with either magnetic field or 
gaseous flow. We also make semiquantitative estimates of grain 
alignment by radiative torques.
We show that in conditions typical of dark cloud interiors, all 
known major mechanisms of grain alignment fail. All the studied
mechanisms predict polarization at least an order of magnitude 
below the currently detectable levels of $\sim 1 \%$. 
On the contrary, in the dark cloud environments where $A_v<1$
the grain alignment can be much more efficient. There the alignment of
suprathermally rotating 
grains with superparamagentic inclusions, and possibly also radiative
torques  account for observed polarization.
These results apply to L1755, which we model in detail and probably also
to B216 and other similar dark clouds.
Our study suggests an  explanation for the difference in 
results obtained through polarimetry of background starlight and
polarized thermal emission from the dust itself. 
We conjecture that the emission polarimetry selectively reveals aligned 
grains in the environment far 
from thermodynamic equilibrium as opposed to starlight polarization
studies
that probe the alignment of grains all the way along the line
of sight, including the interiors of dark quiescent clouds, where no 
alignment is possible.
\end{abstract}


\section{Introduction}

We dedicate this paper to the memory of Edward M. Purcell and Lyman 
Spitzer Jr., 
two pioneers in 
the quantitative study of the interstellar medium.

Discovered nearly half a century ago, alignment of interstellar dust 
(see Hiltner 1949, Hall 1949) has not yet been understood properly. 
Table~\ref{table1} outlines some features of the major known mechanisms
of grain alignment, which are discussed below. To find out 
when a particular mechanism is efficient is a challenging and exciting 
problem to be solved jointly by observers and theorists (see 
Hildebrand \& Dragovan 1995; Desch \& Roberge 1996).  
At present, the lack of a proper understanding of the alignment
processes 
means that it 
is extremely difficult to disentangle variations in polarization
efficiency
from magnetic field variations in polarization maps.

The present research was initiated by recent studies of polarization of
near-infrared background starlight seen through
 dark clouds, namely B216-217 in Taurus
(Goodman et al. 1992) and L1755 in Ophiuchus (Goodman et al. 1995).
In both cases it was shown, that the near infrared polarimetry 
of background starlight 
could not reveal magnetic field structure within these cold dense
regions.
In general, it seems that towards dark clouds polarization does 
increase with extinction  for visual extinction 
$A_{v}<1$ but does not for $A_{v}\gg 1$ (Goodman 1996). A list of
factors,
that can suppress polarization from cloud interiors was presented in
Goodman et al. (1995). Here we analyze only one of those factors,
namely, 
the difference in grain alignment. 
In other words, in our model we assume that grain shape, size and 
composition as well as magnetic field structure are the same in the 
exterior and interior of molecular clouds. 
It is shown in Goodman (1996), that other factors in the  aforementioned 
list can further decrease the efficiency of dichroism for dense 
cloud interiors as compared to diffuse media.

In the present paper we attempt to explain the causes of a
high polarization to extinction ratio for regions with low extinction
($A_v<1$) as opposed to the low ratio of these parameters for quiescent
clouds with high extinction ($A_v\gg 1$). 
Furthermore, we briefly address the challenging problem of why 
thermal emission polarization (see Hildebrandt 1996) observed at far-IR and 
submillimeter wavelengths from regions 
of massive star formation seems to override this tendency.

We address the problem of polarization arising from low extinction and
high
extinction regions using quantitative models of grain alignment
developed 
recently (Draine \& Weingartner 1996 a,b; Lazarian 1995 a,b,c,d;
Lazarian
\& Draine 1997a, Lazarian \& Roberge 1997a,b, 
Roberge, Hanany \& Messinger 1995). 
In particular, we  apply  these theoretical results 
to the filamentary dark cloud L1755 in Ophiuchus. 

In what follows we first provide the data about L1755 that we select
as a representative dark filamentary cloud (Sect.~2),
analyze the dynamics of dust grains (Sect.~3) 
and then calculate the measures of grain alignment in L1755 predicted
by different
mechanisms (Sect.~4). 
The implications of our results for other dense clouds
is given in Sect.~5.

\section{L1755 dark cloud}

\begin{figure}
\begin{picture}(441,270)
\includegraphics{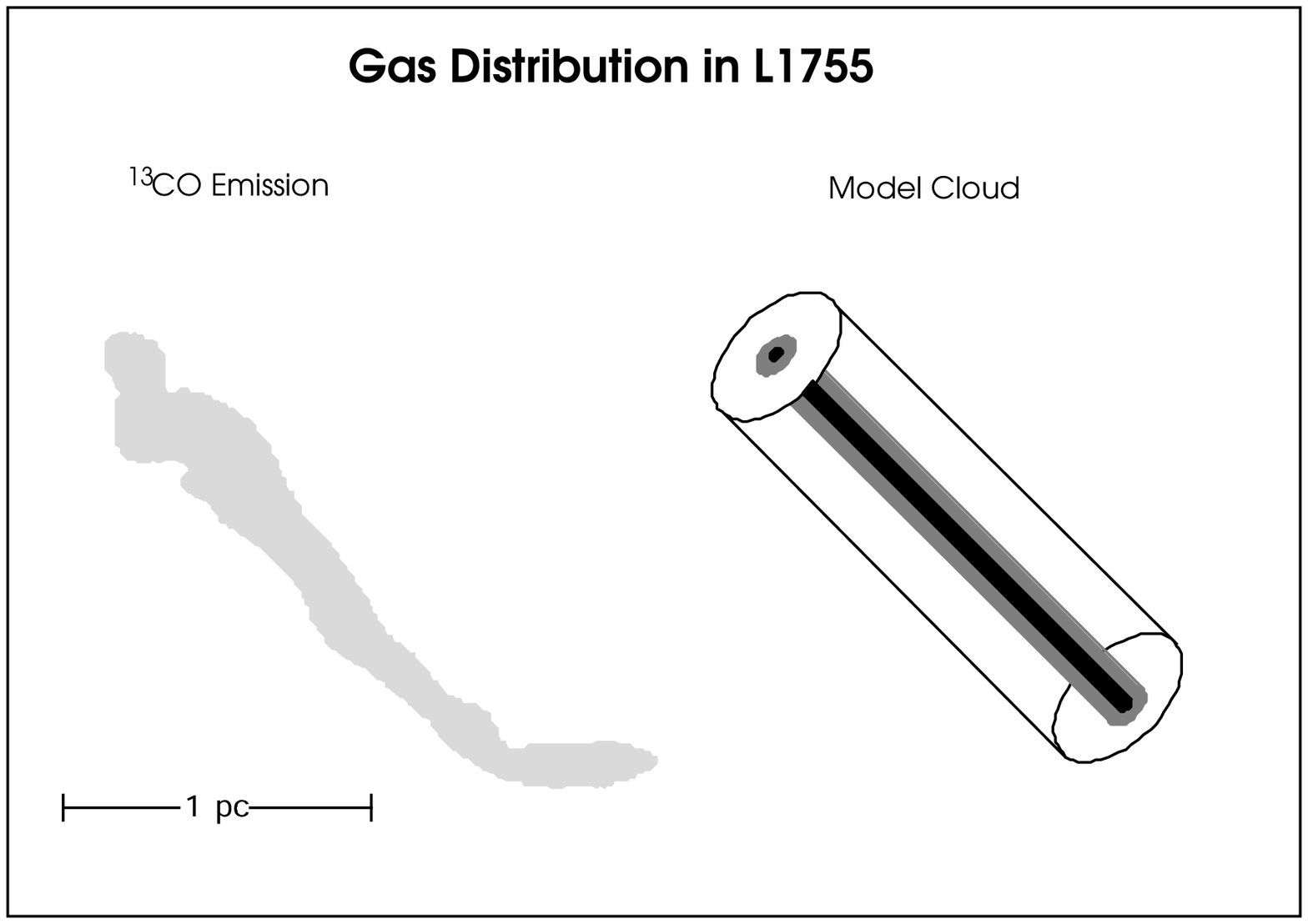}
\end{picture}
\caption{Fig. 1
\label{filament}}
\end{figure}

We select this cloud for
 detailed study because of  the abundance of information about it 
available in
the literature (see Table~2). The extensive studies of $^{13}$CO
in Loren (1989) provide us with the density and velocity information,
the Zeeman studies in Goodman \& Heiles (1994) provide us with
estimates of magnetic field intensity, and optical and near-infrared
polarization data
are available in  Goodman et al. (1990) and in Goodman et al. (1995),
respectively.
An additional advantage is that the extinction of L1755 changes
from $A_v\approx 1$ at the edge to  $A_v\approx 10$ at the center,
which enables us to study the substantial
variations of polarization with extinction
within the same cloud. 
The concentration of atomic hydrogen is taken from the models of
van Dishoeck et al. 
(1993) and Bergin, Langer \& Goldsmith (1995).
 Fractional ionization is 
estimated using the results of Myers \& Khersonsky 
(1995).  The physical parameters in the outermost layers of L1755
are similar to  the neutral diffuse ISM and  they match the 
 parameters of the
dark clouds as density increases (see Table~\ref{table2}). 
In what follows we will call the part  of L1755 within the  cylindrical
radius 
$r_1\approx 0.05$~pc the ``inner region'', the next shell 
up to $r_2\approx 0.15$ pc the ``intermediate region'', and  the regions 
up to $r_3\approx 0.5$ pc the ``outer region''. Such a division involves 
a degree of arbitrariness, but it suits the purpose of obtaining
 crude estimates. The sketch of our 
model of L1755  is presented in Fig.~\ref{filament}. 

In the interior region, the gas density is high ($\sim 10^4$ cm$^{-3}$;
see Benson \& Myers 1989) and the magnetic 
field intensity is  $\approx 30$ $\mu G$. This 
 estimate of field strength corresponds to the
requirement that it lies between values
measured in nearby regions by HI Zeeman observations (Goodman \& Heiles
1994)
 and
the estimates of the field obtained through equipartition arguments
(Myers \& Khersonsky 1995).
For the outer region of the cloud, we assume that
 the field strength decreases to 
match the mean interstellar value $\sim 3$ $\mu G$ (Myers 1987).

The simultaneous change of field strength and gaseous density influences
the velocity with which disturbances can propagate in these magnetized
regions.
 Assuming, that the line broadening is 
caused by Alfv\'{e}nic turbulence (Arons \& Max 1975), we can expect,
that the line width should decrease towards the higher density regions
of
L1755, and this decrease is, in fact, observed.
The ionization ratio should increase outward due to cosmic ray
ionization
as the density decreases and photo-ionization increases as the
extinction
decreases (Myers \&
Khersonsky 1995). 

\section{Grain Rotation}
Before addressing the problem of alignment we should discuss the
processes
that determine grain rotation, since the nature of the rotation is an
important factor in the alignment.

\paragraph{Thermal rotation.}
For thermal 
 rotation it is easy to find
the angular velocity of grain rotation
$\omega_T\approx (kT_{gas})^{1/2}/(\rho a^5)^{1/2}$
where $a$ and $\rho$ are, respectively, the effective ``radius'' of a
grain
and its density, while $T_{gas}$ is the
gas temperature. Thermal rotation is also referred as ``Brownian
rotation''.

\paragraph{Suprathermal rotation.}
It was shown by Purcell (1979) that H$_2$ formation on grain surfaces
spins  grains
up to much larger ``suprathermal'' angular velocities. Indeed, it is
widely believed 
that the formation of molecular hydrogen mostly takes place at
particular catalytic sites, that are frequently called active sites.
From these sites, H$_2$ molecules are being ejected with high
velocities, and as a result the  sites  act as tiny rocket
engines attached to a grain. These ``Purcell rockets'' 
distributed over the grain surface with a mean distance $l$ between them 
can accelerate interstellar grains up to the high angular velocity
$\omega_{H_2}$ (see Table~3)

We assume (see Lazarian 1995c, Roberge, Hanany \& Messinger 1995) 
that in dark clouds the ratio
$\omega_{H_2}/\omega_{T}\approx 1$. The rotation of such grains 
is analogous to the Brownian motion, although the analogy is not exact
(Lazarian \& Draine 1997a).

\paragraph{Rotation arising from radiative torques.}Irregular grains can
be expected to have non-zero helicity and these
 ``helical''
grains scatter left and right circular polarized light differently.
Like the Purcell mechanism, this causes
grain spin-up (Dolginov \& Mytrophanov 1976, 
Lazarian 1995c),\footnote{ Although both scattering of radiation and
$H_2$ formation coerce grains to rotate, we
discuss these causes separately, as the anisotropy of the 
radiation field introduces new effects that we discuss in section 4.}.
A comprehensive study of the effect of these radiative torques 
has started only recently
(Draine \& Weingartner 1996, 1997).  This study has revealed that  
under some conditions
 radiative torques can be stronger than 
``Purcell rocket'' torques. 
Although a full theory of the process
has not yet been completed\footnote{Dolginov \& Mytrophanov (1976)
attempted
to calculate the radiative torques in the Rayleigh-Hans approximation.
However, their predictions were not confirmed by recent numerical runs
in Draine \& Lazarian (1997b). Neither of the studies so far addresses
an important issue of crossovers.}
the encouraging results so far can be treated as 
 preliminary evidence that
grains of irregular shape achieve substantial suprathermal rotation
due to radiative torques.
In Table~\ref{table3} we reproduce  data from numerical runs in
Draine \& Weingartner (1996) to show that the effects of the radiative
torques are substantial both in outer regions of L1755 
and in regions of
massive star formation. Whether Purcell torques or radiative torques
dominate in the outer regions of L1755 is not evident from these data.

\section{Mechanisms of alignment}

\subsection{Davis-Greenstein alignment}

``Davis-Greenstein alignment'' is the paramagnetic alignment of
thermally
rotating grains (see Table 1). A grain whose angular velocity has 
non-zero component $\Omega_{\bot}$ perpendicular to the direction of the
interstellar magnetic field $\bf B$ experiences alternating
magnetization.
This causes dissipation and a corresponding decrease in $\Omega_{\bot}$.
If the time scale for paramagnetic dissipation, $t_r$, is much less
than the time-scale over which gaseous bombardment restores
$\Omega_{\bot}$
(the latter coincides with the gas damping time, $t_d$), grains should
be 
well-aligned in the diffuse ISM. We stress that this condition may
not be satisfied in dense clouds, as the efficiency of paramagnetic
relaxation also depends on the gas-grain temperature ratio (Jones \&
Spitzer 1967) and this ratio is close to unity for dense gas (see Table
2).
Davis-Greenstein alignment also depends on 
the magnetic properties of interstellar grains (see review by Draine
(1996)),
so we need to
distinguish the alignment of ordinary paramagnetic grains and the
alignment
of superparamagnetic and superferromagnetic grains. 
For the sake of  simplicity, in what follows 
we will refer to both casses of
``super''-grains as ``SPM grains''.
  
Theoretical studies of Davis-Greenstein alignment show that this
mechanism
predictes less interstellar 
polarization than observed if grains have
properties of ordinary paramagnetic materials\footnote{Very small grains
may still be aligned, as the paramagenetic relaxation efficiency increases
with the decrease of grain size more quickly 
compared to the efficiency
of randomization. Therfore small grains may trace magnetic field intensity
and the corresponding paper that relates this quantity with the degree of
UV polarization is under preparation (Lazarian \& Martin 1997).} 
(see Mathis 1986, Martin
1995).
Therefore in our study of Davis-Greenstein relaxation 
we assume, that interstellar 
grains are SPM (see experimental evidence
in Goodman \& Whittet 1995).
SPM grains align on a 
time-scale very small 
in comparison to the gaseous damping time (the ratio of the two can
be as small as $10^{-5}$).
As a result, randomization through gas-grain collisions becomes
negligible on the time scale of grain alignment and the measure
of alignment becomes limited only by
the gas-grain temperature ratio.

According to Table~2 $T_{d}/T_{gas}$ is greater than unity for all of
three 
regions of L1755. For these circumstances, Fig.~2 presents a plot of 
the Rayleigh reduction factor (Greenberg 1968) for thermally spinning
SPM grains
\begin{equation}
R=\frac{3}{2}\langle\cos^2\beta-\frac{1}{3}\rangle
\end{equation}
where $\beta$ is the angle between the axis of symmetry of an oblate 
spheroid\footnote{There are indications that aligned grains in molecular 
clouds are oblate rather than prolate (Hildebrand 1988).} and the
ambient 
magnetic field, while here and further on angular brackets
$\langle..\rangle$
denote ensemble averaging. This plot accounts for both incomplete
coupling 
of $\bf J$ with the grain principal axis of maximal inertia 
(henceforth the axis of major inertia) 
and the partial alignment of $\bf J$ with respect to  the external 
magnetic field 
(see Lazarian 1996c). We adopt this $R$ as our measure of
alignment. 

In general, $R$ ranges from $-0.5$ corresponding to  perfect alignment
with
grains longest axis parallel to the magnetic field direction
to $+1$ corresponding to perfect alignment with longer grain axis 
 perpendicular to the magnetic field. $R=0$ corresponds to ``no 
alignment''. Negative values of $R$ in Table 4 and in Fig.~2 indicate
that
grains tend to align with their long axes parallel to the  
external magnetic field and 
this is a consequence of the fact that $T_d>T_{gas}$ (Jones \& Spitzer
1967).
We will call the alignment with negative $R$ ``wrong alignment''
to distinguish it from the ``right alignment'' with $R>0$
that dominates at least in the diffuse interstellar medium. The very
small absolute values of $R$ listed in Table 4 under ``paramagnetic
alignment" for every 
case involving thermal grain rotation  show that Davis-Greenstein
alignment of even SPM grains is
virtually ineffective throughout L1755.

 In specific cases we can relate $R$ to observed  polarization. For 
instance, the intensity of the polarized component of  starlight
passing through a cloud is proportional to the difference in the
maximal and minimal extinction cross-sections for an ensemble of
aligned non-spherical grains. If the extinction cross-section for
radiation with electric vector, ${\bf E}$, parallel to the  
symmetry axis of an oblate grain is $C_{\|}$ and the cross-section
 for ${\bf E}$  perpendicular
to this axis is $C_{\bot}$, then the polarized radiation intensity is
proportional to $R(C_{\bot}-C_{\|})\cos\xi$, where $\xi$ is the angle
that  the magnetic field
makes with the direction of observation (see Roberge 1996).
The difference $(C_{\bot}-C_{\|})$ depends on the optical properties,
oblateness, and sizes of grains (see Draine \& Lee 1984). Even for
perfect
alignment of highly polarising grains
polarization arising from dichroic extinction
does not exceed a few percent. To account for the observed polarization 
$\sim 1\%$, $R>0.1$ is usually required (see Whittet 1992).

No  appreciable Davis-Greenstein  alignment is expected
anywhere in L1755 
 where gas and dust temperatures are nearly equal. 
 This conclusion is {\it a fortiori } true if grains are not
superparamagnetic. (See Table 4.)

\subsection{Purcell alignment}

Purcell alignment is the paramagnetic alignment of grains rotating
suprathermally. Such grains are not susceptible to disorientation
arising from gaseous bombardment. As a result, Purcell alignment
does not depend on the gas-grain temperature ratio and (2)  even 
if they are not superparamagnetic  grains
can be efficiently aligned in the typical interstellar magnetic 
field (Lazarian \& Draine 1997, Draine \& Lazarian 1997) (see Table~1).

As shown above, the inefficiency of Davis-Greenstein alignment 
of SPM grains in L1755 
stems from the insignificant difference in 
grain-gas temperatures. However, it was pointed out in Sec.~3 
that grain rotation can be 
substantially faster than the gas kinetic temperature, 
and that ``Purcell rockets''
may be a major cause of suprathermal rotation. 
For these rockets to work,
 a supply of ``fuel'' is required. This fuel is atomic
hydrogen in the ambient gas. As the concentration of atomic hydrogen
decreases 
towards the inner regions of L1755 grain rotation
slows down  (see Table~\ref{table2}) . However, in the outer regions of
L1755 the concentration 
of atomic hydrogen approaches that of diffuse regions and the rotation 
is suprathermal (see Table~3). 
Therefore,  according to the results of
 Spitzer \& McGlynn (1979; see also Table 4), those grains which are
SPM should be perfectly aligned in 
the outer regions of L1755. 

A different question is the alignment of suprathermally rotating 
grains of ordinary paramagnetic
substance. 
Until recently,
it was believed that an appreciable Purcell alignment can only be
achieved under a rather restrictive condition of the long-lived spin-up
(Spitzer \& McGlynn 1979, Lazarian 1995b). However
a study by Lazarian \& Draine (1997a) has revealed that a substantial
alignment is attainable even for short-lived spin-ups. This alignment
arises from a high degree of correlation between grain angular momentum
$\bf J$ before and after a crossover. This correlation was missed by
earlier research which assumed that during spin-ups
$\bf J$ is perfectly aligned with
the grain axis of major inertia (see Spitzer \& McGlynn 1979). 
Recent study of
this internal alignment (Lazarian\& Roberge 1997a) has shown
that coupling of $\bf J$ with  the major inertia axis 
is valid up to thermal fluctuations
within the grain material. These fluctuations, contrary to
intuitive expectations, enhance the alignment.
Our estimates that account for thermal fluctuations (see Table~4)
indicate that the Purcell mechanism can provide notable 
alignment of ordinary paramagnetic grains (i.e. $R\approx 0.7$) 
in the outer region of L1755 
 (see Table~4).

For intermediate regions in Fig.~1 where $\omega_{H_2} <
\omega_{T}$, the effective grain rotational temperature will be at most
twice
that of the ambient gas. Such rotation is only marginally suprathermal
and disorientation during spin-ups is substantial. Even under the most 
favorable conditions $R$ is still $<0.1$ in the intermediate region.

The Purcell alignment is negligible for the inner region of L1755 
(see Table~4).

\subsection{Radiative torques}

As pointed out in section 2, radiative torques are different from
those arising from H$_2$ formation. Unlike the isotropic bath of H atoms
that 
provide the ``fuel''
for Purcell's spin-up, the photons needed to drive radiative alignment
are often anisotropically distributed in space.
This anisotropy results in a peculiar mechanism of 
grain alignment (see mechanism 5 in Table 1), which tends to align the
short
axis of the grain with the
magnetic field (i.e.``right alignment").

The matter of radiative torques is currently an issue of an intense 
theoretical research
(Draine \& Lazarian 1997), 
and a number of questions need to be answered before quantitative 
conclusions are to be drawn. Therefore some degree of uncertainty 
is present in our discussion here. Our conclusions here are based on
the results in Draine \& Weingartner (1996, 1997). A study in
Draine \& Lazarian (1997) has shown that for generally accepted values
of active sites density and H$_2$ formation efficiency radiative
torques only marginally alter the dynamics which is dominated by
H$_2$ torques. However, in view of huge uncertainties involved,
it is also possible that the alignment by radiative torques may
be the dominant mechanism of alignment. We hope that future theoretical
work in the field will provide us with detailed predictions that can be 
used for testing the theory against observations.

Table~4  shows that for both the intermediate and inner regions of 
L1755 the radiative 
torques are too weak to produce any measurable
 grain alignment. At the same time, these torques can be
important in the outer regions of L1755.

\subsection{Mechanical alignment}

Both thermally and suprathermally rotating grains, can be aligned due 
to  collisions with a gaseous flow. 
The mechanical alignment of thermally rotating grains was described by 
Gold (1951), while  the 
mechanical alignment of {\it suprathermally} rotating grains was
described
only recently
(Lazarian 1995a; see Table 1).

The important factor in mechanical alignment is supersonic gas-grain 
drift. It was pointed out by Lazarian (1994) that Alfv\'{e}n  waves 
can cause such drift and this entails alignment with grain long axes
perpendicular to magnetic field lines. i.e. the ``right alignment''.
 The formal theory of grain drift is given in Lazarian 
\& Draine (1997b) and alignment measures are calculated in
 Lazarian (1994, 1995a, 1997a), Lazarian \& Efroimsky (1996), and
Lazarian, Efroimsky \& Ozik (1996).

A spectrum of Alfv\'{e}nic waves should be present in the interstellar 
medium, and  Alfv\'{e}nic velocities are greater than sound 
velocities for most interstellar conditions (Myers 1987). 
In these circumstances, a 
crucial test of whether Alfv\'{e}nic waves cause supersonic drift is 
to compare the Larmor frequency of grain gyration about magnetic 
field lines with the cut-off frequency of Alfv\'{e}nic spectrum 
(Lazarian 1994, Lazarian \& Draine 1997b). 
This comparison is difficult because the measured quantities needed to 
estimate the Larmor frequency  are uncertain. 

If grains are collisionally charged, then in regions with sufficent
ionisation,
 Alfv\'{e}n waves can cause supersonic grain drift when the grain Larmor
frequency is lower
than the cut-off frequency for the waves, 
\begin{equation}
\omega_{max}\approx 1.2 \times
10^{-10}\left(\frac{n_{gas}}{10^{3}}\right)
\left(\frac{x_{e}}{4\times 10^{-5}}\right)~~~
 ~{\rm s}^{-1},
\label{a2}
\end{equation}
where  $x_{e}$ is the ionization ratio. 

For the outer region of L1755, $\omega_{max}$ is of the order of 
$10^{-9}$~s$^{-1}$, and the grains there may experience supersonic
drift.
For grains in the intermediate and inner 
regions, $\omega_{\rm max}$ becomes $\sim 6\cdot 10^{-12}$~s$^{-1}$ 
and 3$\cdot10^{-14}$~s$^{-1}$, respectively. Therefore supersonic drift
is 
only marginally 
possible for  the intermediate region and impossible for inner regions
of 
L1755.
In Table~4 we provide an estimate for grain alignment in the
intermediate 
region, assuming a collisional charging of the grains. One source of
uncertainty in these predictions is a possible helicity of grains 
(Lazarian 1995b). Unfortunately a quantitative theory of such an
alignment
is still to be developed. 
 
If grains are more highly charged 
through photoemission (Draine 1978) in any region, then the Larmor
frequency is much greater
than $\omega_{max}$ and supersonic motion is impossible. This can
suppresses
the alignment by Alfven waves.

In the presence of supersonic ambipolar diffusion, both thermally and
suprathermally rotating grains can be aligned mechanically (see Table
4).
However, we do not have evidence for supersonic ambipolar diffusion
in L1755. Therefore we do not discuss this possibility here.

\subsection{Exotic mechanisms of alignment}

One mechanism omitted from Table~1 can still 
 be efficient in the interior regions of L1755. 
This mechanism,
 suggested by Spitzer and Tukey (1951),
is based on the alignment of magnetic moments of ferromagnetic grains 
and has been shown to be very inefficient in the diffuse clouds (Davis
\& Greenstein 1951). 
However, its efficiency scales as $\frac{B}{T_{gas}}$, and therefore
if grains are ferromagnetic, it may be active in dark clouds, where the
ambient magnetic field $B$ is
higher than in diffuse clouds, while the temperature is
lower (see Table~2). For a population of exclusively ferromagnetic
grains, the measure of 
alignment is expected to be of the order of few per cent (see Lazarian
1995c), but if the 
abundance of such grains is  negligible this mechanism is of
marginal importance for L1755. This mechanism will produce ``wrong
alignment''
with the long grain axis parallel to $\bf B$.

It has been suggested that since shocks can dissociate H$_2$, 
Purcell alignment might be enhanced in shocked regions of
cloud interiors (Johnson 1982).
Shocks also cause ambipolar diffusion which may produce alignment 
(Roberge \& Hanany 1991, Lazarian 1994, Roberge et al. 1995). 
The presence of 
intense sources of cosmic rays or X-rays in the vicinity of a cloud 
 may induce grain alignment. 
Such radiation can excite grain rotation,  enabling 
 Davis-Greenstein alignment of SPM grains 
(Lazarian \& Roberge 1997b) and/or provide conditions for alignment 
by Alfv\'{e}nic waves (Lazarian 1994, 1995, Lazarian \& Draine 1997b).
As none of these special conditions  are especially 
 applicable to L1755, we conclude
that {\it all} the known major mechanisms fail in the cloud interior.

\section{Discussion}

\subsection{Joint Action of several mechanisms}

We have seen above that all the major mechanisms (see Table~1) 
fail to produce grain 
alignment in the interior and intermediate regions of L1755. 
On the other hand, grain alignment in the outer regions of L1755 
is efficient. If, for example,  both (1) the Purcell mechanism, providing
alignment with measure $R_1$, and 
 and (2) mechanical alignment by Alfv\'{e}nic waves, providing the
alignment with the measure $R_2$,  act simultaneously
the overall alignment measure can be calculated using the approach 
suggested in Lazarian (1996). For suprathermally rotating grains the 
formula for the Rayleigh reduction factor is 
\begin{equation}
R\approx \frac{R_1+R_1R_2+R_2}{1+2R_1R_2}
\end{equation}
and this provides $R> 0.8$ for the outer region of L1755 in our example.
As we pointed out above, the suprathermal rotation of grains arises
in outer parts of L1755 as the result of H$_2$ formation on grain
surfaces.

The measure of
grain alignment 
versus grain size should be different for radiative torques and 
superparamagnetic relaxation. This provides a way to distinguish the 
mechanisms. 
Another method of separating  the mechanisms  is to study how alignment 
varies with environmental conditions. For instance, paramagnetic 
alignment depends specifically on the concentration of atomic hydrogen,
while 
the alignment caused by radiative torques varies only with extinction. 
Table~3 indicates that at some optical depth 
Purcell alignment continues, while the alignment by radiative torques
may fail.

\paragraph{Extension of the theories to emission polarimetry.} 

The results of studies of polarized thermal emission in the 
submillimeter 
and far-infrared differ substantially from the studies of background
starlight
polarization in near infrared and optical range
(compare M17 polarization maps in Goodman 1996 and Dotson 1996). 
The near infrared background starlight polarimetry indicates that 
 grains are not aligned for $A_v\gg 1$, but the far-infrared emission 
polarimetry reveals aligned grains within clouds of high opacity. 

From the point of view of grain alignment theory this paradox 
can be explained as a selection effect. Indeed, far-infrared emission 
 polarimetry 
 selects warm grains, often
 in the vicinity of young massive stars. Such grains 
find themselves in the environments far from equilibrium. For instance, 
our Table 3 shows that these grains will rotate suprathermally due to 
radiative torques. On the contrary, the near infrared 
background starlight observations  are 
sensitive to the whole sample of grains along the line from the source
to the 
observer. The situation is aggravated by the fact that 
observations of background starlight in the optical and near infrared
are
most often performed in dark quiescent clouds, while the observations 
of 
far-infrared emission  are most often
 performed in the regions of massive 
star formation (see Davidson et al. 1995, Dotson 1995, Hildebrand \&
Dragovan 
1995). This means direct comparison of background starlight
polarimetry and thermal emission polarimetry is diffucult using existing
data.

In dark clouds like L1755, 
we expect to detect  polarized emission emanating from the 
vicinity of any low mass stars formed there, but the bulk of the cloud 
interior should not emit much polarized radiation if 
our understanding of the conditions within L1755 and our theory of grain 
alignment are adequate. 

\paragraph{Magnetic fields from polarization maps.}
L1755 is an ordinary filamentary 
dark cloud and therefore our conclusions should 
be applicable to the whole class of similar objects. Thus we expect 
near-infrared 
background starlight 
polarimetry to reveal the structure of magnetic fields only in dense 
cloud {\it exteriors}, and to have no relation to the field structure
for the 
regions corresponding to $A_v\gg 1$. To 
what particular extinction magnetic fields can be traced using 
background star light polarimetry remains to be determined (see Arce et
al. 1997). This important 
question should be addressed  by concerted efforts of 
observers and theoreticians. We may further expect that emission
polarimetry 
studies of the regions similar to L1755 may not give insight into the 
magnetic field structure for these cold high extinction regions, apart 
from small regions surrounding low-mass stars  forming there. 

On the contrary, emission studies of regions of massive star formation 
should 
be 
a reliable tool for revealing magnetic field structure
there.\footnote{Although
 winds and outflows in these 
regions may cause mechanical alignment with grain long axes parallel to 
magnetic field lines, results in Lazarian (1997a) shows that 
such an alignment is an exception rather than a rule.} This technique,
however, is biased against cold regions where dust emission is
low.

\section{Conclusions}

Our quantitative study of six types of alignment mechanisms predicts
almost no polarization from the interiors of dark clouds and appreciable
polarization from the cloud exteriors. Adopting L1755 as a test case, we
have shown that a number of 
mechanisms, including
paramagnetic alignment of
 SPM grains  and alignment by radiative torques,  
provide efficient orientation of grains 
in the outermost regions of this cloud. At some moderate optical depth, 
still in the outer region, the Purcell
alignment of ordinary paramagnetic grains may  dominate radiative
torques. Because all the alignment mechanisms studied
 fail in the interior region of L1755 we do not expect significant
alignment there. This difference in alignment efficiency
 can provide 
a natural explanation of why no increase of polarization with optical 
depth is seen in L1755 (Goodman et al. 1995), and it is consistent with
a
prediction 
of only marginal polarization of thermal emission emanating 
from grains in the interior of L1755. 

To test the picture presented here it might be useful to measure
new-infrared
polarization through a dark cloud which contains a cold dense inner
zone, as in L1755, and also a localized heat sourse, such as an embedded 
cluster.

\acknowledgements 
A. L. gratefully acknowledges the support of NASA grants  NAG5 2773 
NAG5 2858 at Princeton University 
and of the Visiting Fellowship at the Harvard-Smithsonian Center for 
Astrophysics. 

\pagebreak

\def\tempa{(1) Gold}
\def\tempb{Gold 1952}
\def\tempc{alignment of thermally rotating grains aligned by supersonic
flows;
  {\it originally}: 
  radiation pressure on grain; {\it further development}: 
  Alfv\'{e}nic waves  (Lazarian 1994, Lazarian \protect\& Draine 1997b) 
  ambipolar~diffusion (Roberge et al. 1995).}
\def\tempe{\mbox{Purcell 1969}, \mbox{Purcell \& Spitzer} 1971, Dolginov
\& 
 Mytrophanov 1976, \mbox{Lazarian 1994}, \mbox{Roberge et al.} 1995,
\mbox{Roberge \& Hanany} 1990
 \mbox{Lazarian 1997a}}
\def\tempf{Supersonic drift, rotation with thermal velocities} 
\def\tempba{(2) Mechanical alignment of suprathermally rotating grains}
\def\tempbb{Lazarian 1995a}
\def\tempbc{alignment by suprathermally rotating grains by supersonic
flows
  due to cross-section difference and during due to gaseous bombardment
during  cross-over  events}
\def\tempbe{\mbox{Lazarian 1995a,c}, Lazarian \& Efroimsky 1996,
Lazarian, Efroimsky \& Ozik 1996}
\def\tempbf{Supersonic drift, rotation with suprathermal velocities}
\def\tempca{(3) Davis-Greenstein  }
\def\tempcb{Davis-Greenstein 1951}
\def\tempcc{alignment of thermally rotating grains through 
  paramagnetic relaxation;
 {\it originally}: relaxation of paramagnetic grains;
 {\it further development}: relaxation of SPM
grains (Jones \& Spitzer  1967).}
\def\tempce{\mbox{Jones \& Spitzer} \mbox{1967, Purcell \&} 
 \mbox{Spitzer 1971},  \mbox{Mathis 1986}, \mbox{Roberge et al.} 
 \mbox{1993,  Lazarian} \mbox{1995d, Lazarian }
  \& Roberge 1997a, \mbox{Lazarian 1997b}}
\def\tempcf{Presence of SPM impurities}
\def\tempda{(4) Purcell}
\def\tempdb{ Purcell 1979, Spitzer \& McGlynn 1979}
\def\tempdc{alignment of suprathermally rotating grains through
paramagnetic 
  relaxation; {\it originally}: efficiency of alignment is limited
for ordinary paramagnetic grains (Spitzer \& McGlynn 1979);
  {\it further development}:  
incomplete Barnet relaxation enhances alignment (Lazarian
\& Draine 1997a)}
\def\tempde{\mbox{Purcell 1979}, Spitzer \& McGlynn 1979, Lazarian
1995c,e,
Lazarian \& Draine 1997a, Draine \& Lazarian 1997a}
\def\tempdf{Supra\-thermal rotation due to H$_2$ formation}
\def\tempea{(5) Alignment by radiation torques}
\def\tempeb{Draine \& Weingartner 1996, 1997}
\def\tempec{alignment due to the difference in scattering right 
  and left polarized quanta}
\def\tempee{Draine \& Weingartner 1996,1997; Draine \& Lazarian 1996 }
\def\tempef{Radiation of short wavelength}
\def\tempfa{(6) Mechanical alignment of helical grains}
\def\tempfb{Lazarian 1995b }
\def\tempfc{helical grains aligned by supersonic flows; 
 atoms bounce off the grain surface of helical 
  shape or off  a grain with \mbox{variation} of the accommodation
coefficient}
\def\tempfe{does not exist}
\def\tempff{Supersonic drift, special shape}

\begin{table}
\begin{tabular}{|p{2.5cm}|p{2cm}|p{6.5cm}|p{3.5cm}|p{2cm}|}
\hline
 & & & &\\
Alignment mechanism    & Introduced  & {\hspace{2cm} Description} 
  &  Quantitative theory & Special conditions for success \\
 & & & &\\
\hline
 & & & &\\
\tempa  & \tempb  & \tempc  & \tempe  & \tempf\\
 & & & &\\
\hline
 & & & &\\
\tempba & \tempbb & \tempbc & \tempbe & \tempbf\\ 
 & & & &\\
\hline
\end{tabular}
\end{table}

\begin{table}
\begin{tabular}{|p{2.5cm}|p{2cm}|p{6.5cm}|p{3.5cm}|p{2cm}|}
\hline\\
Alignment mechanism    & Introduced  &  {\hspace{2cm}Description }
  &  Quantitative theory & 
Special conditions for success \\ 
 & & & &\\
\hline
 & & & &\\
\tempca & \tempcb & \tempcc & \tempce & \tempcf\\ 
 & & & &\\
\hline
 & & & &\\
\tempda & \tempdb & \tempdc & \tempde & \tempdf\\
 & & & &\\
\hline
 & & & &\\
\tempea & \tempeb & \tempec & \tempee & \tempef\\
 & & & &\\
\hline
 & & & &\\
\tempfa & \tempfb & \tempfc & \tempfe & \tempff\\
 & & & &\\
\hline
\end{tabular}
\caption{Major mechanisms of grain alignment.}
\label{table1}
\end{table}

\pagebreak

\begin{table}
\begin{center}
\begin{tabular}{|c|c|c|c|}
\hline
                          & $r<0.05$ pc     & $0.05\leq r<0.15$ pc 
                                                              &
$0.15<r\leq 0.5$ pc     \\
\hline
$T_{gas}$ (K)             &  10             & 12              & 
18          \\
\hline
$T_{d}$  (K)              &  12             & 20              & 
20          \\
\hline
$n_{gas}$ (cm$^{-3}$)     &  $3\cdot 10^{4}$       & $3\cdot
10^{3}$         
							      & $3\cdot 10^{2}$ \\
\hline
B($\mu$G)                 & 30              & 20              & 
10          \\
\hline
$\sigma_{v}$ (km s$^{-1}$)& 0.2             & 0.4             &  
0.6          \\
\hline
$x_{e}$                   &$2\cdot 10^{-7}$&$6\cdot 10^{-6}$
                                                              &
$1.3\cdot 10^{-4}$    \\
\hline
$n_H/n_{gas}$            &$\sim3\cdot10^{-3}$
			  		   & $\sim10^{-2}$
							      &$\sim 10^{-1}$   \\
\hline
\end{tabular}
\end{center}
\caption{Physical quantities relevant to models of grain alignment in
L1755
in Ophiuchus. We used data from Benson \& Myers (1989),  
Bergin, Langer \& Goldsmith (1995), 
de Geus, Bronfman \& Thaddeus (1990),  Goodman et al. (1995), 
 Goodman \& Heiles (1994), Loren (1989a,b), 
McKee (1989), Myers \& Khersonsky (1995),  Wood, Myers \& Daugherty
(1994) 
and van Dishock et al. (1993). The gas temperature $T_{gas}$is taken 
from NH$_{3}$ and CO$(12)$ lines; each uncertain by $\sim 2$K. 
For inner region, $T_{gas}$ approaches 
the dust temperature $T_d$, which corresponds to the notion of
thermal coupling of gas and grains in molecular clouds. 
The value of $T_d$ for outer regions follows from
estimates for standard radiation field (Draine 1978). 
The values of gas density $n_{gas}$ 
are estimated from NH$_{3}$ and CO$_{(13)}$ lines and the requirement
that $A_{v}$ increases towards the cloud center. The estimate for
magnetic field $B$ follows from the requirement that it lies between (1)
the ``background'' measured in the nearby regions in HI Zeeman (Goodman
\& 
Heiles 1994) and (2) the ``equipartition'' field strength (Myers \&
Khersonsky
1995). The estimate of the velocity dispersion $\sigma_{v}$ is based on 
$\sigma_v=FWHM$ of line profile =$(8\ln 2)^{1/2} \sigma_{v}$ for 
NH$_{3}$ and $^{13}$CO lines. The ionization ratio $x_{e}=n_{e}/n_{gas}$
is based on a model of photoionization and cosmic ray 
ionization for a clumpy cloud (Myers \& Khersonsky 1995).}
\label{table2}
\end{table}

\begin{table}
\begin{center}
\begin{tabular}{|c|c|c|c|c|c|c|}\hline
             &\multicolumn{3}{|c|}{Outer region of L1755}
             &\multicolumn{3}{|c|}{Dark cloud interiors}\\
\hline

&\multicolumn{3}{|c|}{
 $T_{gas}=18$~K, $T_d=20$~K,
 $n_H=30$ cm$^{-3}$,}             
			& \multicolumn{3}{|c|}{
	                 $T_{gas}=10$~K, $T_d=12$~K, $n_H=10$ cm$^{-3}$, } \\
&\multicolumn{3}{|c|}{
 $u_{rad}=u_{ISRF}$,
 $n_{gas}=3\cdot 10^2$~cm$^{-3}$}
			& \multicolumn{3}{|c|}{
		$u_{rad}=0.07u_{ISRF}$,  $n_{gas}=3\cdot10^4$~cm$^{-3}$}\\

\hline
              &$\omega_T$, s$^{-1}$&$\omega_{rad}^2/\omega_T^2$ & 
                                                  $\omega_{\rm H_2}^2/
                                                        \omega_T^2$ & 
               $\omega_T$, s$^{-1}$&$\omega_{rad}^2/\omega_T^2$ & 
                                                  $\omega_{\rm H_2}^2/
                                                        \omega_T^2$ \\
\hline
$a=0.02~\mu$m &$3.1\cdot 10^6$ &$1.2\cdot 10^{-5}$ & $8.4\cdot 10^2$&
               $2.2\cdot 10^6$   &$ 1.8\cdot 10^{-6}$ & U\\
\hline
$a=0.05~\mu$m &$3.2\cdot 10^5$ &$2.4\cdot 10^{-2}$  & 36 &
               $2.3\cdot 10^5$ &$7.0\cdot 10^{-5}$ & U            \\
\hline
$a=0.2~\mu$m  &$9.9\cdot 10^3$ &$22 $    & $158$ &
               $7.0\cdot 10^3$ &0.1               & U  \\
\hline
\end{tabular}
\end{center}

\begin{center}
\begin{tabular}{|c|c|c|c|}\hline
             &\multicolumn{3}{|c|}{Clouds forming massive stars}\\
\hline
 &\multicolumn{3}{|c|}{$T_{gas}=45$~K, $T_d=45$~K,
$n_H=10^3$~cm$^{-3}$,}\\
 &\multicolumn{3}{|c|}{$u_{rad}=240u_{ISRF}$,  $n_{gas}=10^5$~cm$^{-3}$
$n_H=10^3$~cm$^{-3}$ }\\
\hline
             &$\omega_T$, s$^-1$ &$\omega_{rad}^2/\omega_T^2$ & 
                                                  $\omega_{\rm H_2}^2/
                                                        \omega_T^2$ \\
\hline
$a=0.02~\mu$m  &$4.7\cdot 10^6$ &$1.4\cdot 10^{-2}$ & U  \\
\hline
$a=0.05~\mu$m  &$4.7\cdot 10^5$ &1.3                & U \\
\hline
$a=0.2~\mu$m   &$1.5\cdot 10^4$ &$1.4\cdot 10^3$    & U  \\
\hline
\end{tabular}
\end{center}
\caption{ Rotational velocities of grains in diffuse interstellar media, 
dark clouds and regions of massive star formation. $\omega_{rad}$ 
taken from Draine \& Weingartner (1996) do not account for the 
starlight reddening (the latter decreases radiative torques).
We assumed the
accommodation coefficient $\gamma$ equal 0.1. The symbol
`` U'' corresponds 
to ``uncertain'' and reflects the existing theoretical ambiguity in 
determining the number of active sites when the photodesorption is low. 
$u_{ISRF}\approx 8.6\cdot 10^{-13}$ erg$\cdot$cm$^{-3}$
corresponds to the energy of the 
averaged interstellar radiation field (Draine 1978).}  
\label{table3}
\end{table}

\begin{table}
\begin{center}
\begin{tabular}{|c|c|c|c|c|}\hline
\multicolumn{5}{|c|}{outer region} \\
\multicolumn{5}{|c|}{$A_v<0.9$ mag, $0.15$ pc $<r<0.5$~pc}  \\
\hline 
              & mech. &\multicolumn{2}{|c|}{param.}       & radiative
torques \\
\cline{3-4}
              &       &ordinary        & inclusions       & \\
\hline
thermal       &	 0.08 &$-9\cdot10^{-6}$ &  $-1.6\cdot 10^{-3}$ 	  & NA 
\\
              &   (1) &   (3)          &  (3)             &   \\
\hline
suprathermal   &$0.22$ &$0.7$          & $ {\bf 1}$      & ${\bf \sim
1}$\\
              & (2)       &  (4)       &  (4)             & (5)\\
\hline
\end{tabular}
\end{center}

\begin{center}
\begin{tabular}{|c|c|c|c|c|}\hline
\multicolumn{5}{|c|}{intermediate region} \\
\multicolumn{5}{|c|}{$A_v\approx 3$ mag, $0.05$~pc$<r<0.15$~pc} \\
\hline 
              & mech. &\multicolumn{2}{|c|}{param.}       & radiative
torques \\
\cline{3-4}
              &       &ordinary        & inclusions       & \\
\hline
thermal       &	0.08  &$-2.4\cdot 10^{-5}$ & $-5.7\cdot 10^{-3}$ & NA \\
	      & (1)   & (3)            & (3)              & \\
\hline
suprathermal   & {\bf 0.1}     &  $7\cdot 10^{-5}$ & {\bf 0.1}     &
$\sim 0$\\
	      &  (2)     & (4)        &   (4)               & (5) \\
\hline
\end{tabular}
\end{center}

\begin{center}
\begin{tabular}{|c|c|c|c|c|}\hline
\multicolumn{5}{|c|}{inner region} \\
\multicolumn{5}{|c|}{$A_v> 3$ mag, $r<0.05$~pc} \\
\hline 
              & mech. &\multicolumn{2}{|c|}{param.}       & radiative
torques \\
\cline{3-4}
              &       &ordinary        & inclusions       & \\
\hline
thermal       &	0     &$-1.7\cdot 10^{-5}$ & $-3\cdot 10^{-3}$ & NA \\
	      & (1)   & (3)            & (3)              & \\
\hline
suprathermal  & 0     &$6\cdot 10^{-5}$&  {\bf 0.01}           & 0\\
	      &(2)    & (4)        &   (4)               & (5) \\
\hline
\end{tabular}
\end{center}
\caption{Predicted grain alignment measure $R$ in regions of
L1755.
 The grain axis ratio adopted is $2/3$ and grain radius is $10^{-5}$
cm.  The notation NA correspond to ``not applicable''. 
The calculations are made using results from Draine \& Weingartner 1996,
Lazarian 1994, 1995c, 1996, Lazarian \& Draine 
1997a, Lazarian \& Roberge 1997a. The number for $R$ arising from
radiative
torques is still uncertain and $R\approx 1$ is our educated guess.
The numbers in round brackets 
corresponds to the number of the mechanism in Table 1. 
The estimates for non-thermal
alignment in the intermediate region are obtained assuming that
$\omega_{H_2}/\omega_{T}\sim 1$ (see Table 3). In this case $R$ is
positive
as the effective temperature of grain rotation is greater than the
temperature
of grain material. Bold letters correspond to the measures of alignment
that dominate. For the inner region, in the presence of supersonic 
ambipolar diffusion the measures of mechanical alignment would coincide 
with the corresponding column of the intermediate region table.}
\label{table4}
\end{table}

\clearpage
{\bf Figure Captions}

{\bf Fig.1}\\
Region with intensity $> 3$K $J=1$--$ 0$ line of $^{13}$CO in 
L1755 (Loren 1989) and the model adopted. 
The interior region has radius (shown dark) $r_1= 0.05$~pc, 
the intemediate region has radius $r_2= 0.15$~pc, 
and the outer region has radius $r_3= 0.5$~pc.

{\bf Fig.2}\\
The measure of Davis-Greenstein alignment for superparamagnetic
grains as a function of $y=T_d/T_m$, $T_m=(T_{gas}+T_d)/2$ for   
the grains with the ratio of moments of inertia $I_z/I_{\bot}=8/5$.
The negative measures of alignment correspond to alignment with long
axes preferentially directed along magnetic field lines. This is
the concequence of $T_{d}$ being greater than $T_{gas}$.

\begin{figure}
\begin{picture}(441,270)
\includegraphics{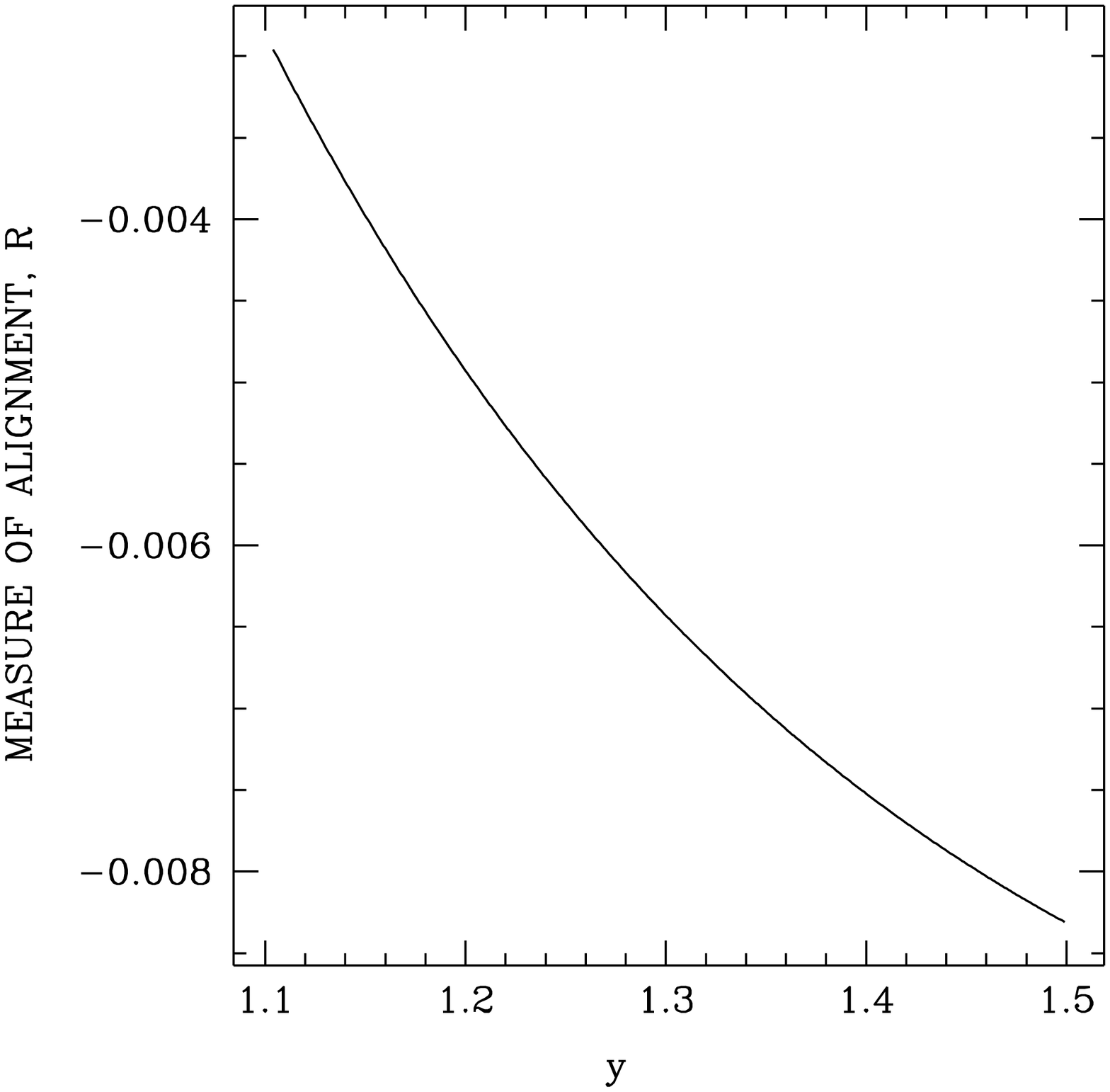}
\end{picture}
\caption{ Fig. 2
\label{vect}}
\end{figure}

\end{document}